%% file: SpinningSuperparticle.tex
\documentclass[a4paper,11pt]{article}
\pdfoutput=1 

\usepackage{jheppub} 
\usepackage[T1]{fontenc} 

\usepackage{natbib}
\usepackage{physics}
\usepackage{tensor}
\usepackage{easybmat}
\usepackage{bbm}

\usepackage{mathtools}
\def\multiset#1#2{\ensuremath{\left(\kern-.3em\left(\genfrac{}{}{0pt}{}{#1}{#2}\right)\kern-.3em\right)}}

\usepackage{tikzit}
\input{sample.tikzstyles}

\preprint{\vbox{\hbox{\hphantom{XXX}LMU-ASC 09/22}}}
\title{\boldmath Dark Energy and the Spinning Superparticle}

\author{Daniel Bockisch}
\author{\& Ivo Sachs}

\affiliation{Arnold-Sommerfeld-Center for Theoretical Physics, Ludwig-Maximilians-Universität M\"unchen,\\Theresienstr. 37, D-80333 München, Germany}

\emailAdd{Daniel.Bockisch@physik.uni-muenchen.de}
\emailAdd{ivo.sachs@physik.lmu.de}

\abstract{We revisit the theory of background fields constructed on the BRST-algebra of a spinning particle with $\mathcal{N}=4$ worldline supersymmetry,  whose spectrum contains the graviton but no other fields.  On a generic background, the closure of the BRST-algebra implies the vacuum Einstein equations with a cosmological constant that is undetermined. On the other hand, in the "vacuum" background with no metric, the cohomology is given by a collection of free scalar- and vector fields. Only certain combinations of linear excitations, necessarily involving a vector field, can be extended beyond the linear level with the vector field inducing an Einstein metric.  
}

\begin{document} 
\maketitle
\flushbottom

\section{Introduction}
\label{sec:intro}

 Suppose that there exists a civilization on a planet with a completely opaque atmosphere, such as Venus, having nevertheless  managed to derive the laws of perturbative quantum field theory, in particular BRST quantization and have independently developed Riemannian geometry. However, due to the opaque atmosphere, they have so far stayed completely oblivious to the structure of the Universe and the laws that govern it, since no large-scale space-time observations via light-based astronomy could have been made. Based on this hypothetical situation, in this note we follow a purely particle physicists approach to the geometry of space-time. 

The first question we address in this note is: Could they nevertheless make predictions about the nature if gravity with the tools at their disposal? In particular, assuming the existence of the graviton, could they infer from it the existence of any kind of non-linear theory of gravity. This has been found to be the case long time ago (see \cite{Deser:1969wk} and references therein). Here we present an alternative argument by imposing that BRST-cohomology in any physical metric background is that of the graviton with two degrees of freedom. 
This leads to the conclusion, already implied in \cite{Bonezzi:2018box}, that the metric background has to be an Einstein space, that is, in the absence of sources we have $R_{\mu\nu}=\lambda g_{\mu\nu}$ with $\lambda$ a constant. This arises not from a variational principle of some action but simply from the existence of a linearized graviton. 

Having obtained the non-linear Einstein equation for the metric background our friends from Venus, being particle physicists, may then wonder what the vacuum configuration of this theory is, that is, no metric. While this limit is difficult to treat in Einstein's theory it appears to be perfectly natural in the BRST formulation that we will describe. The cohomology of the BRST charge $Q_0$ in the vacuum background contains a collection of scalar fields and vector fields, but no graviton, of course, since the latter corresponds to a small variation of a background metric. Just as we did for the Minkowski background, we then consider non-linear deformations $\delta Q$ of the vacuum BRST charge $Q_0$ by background fields corresponding to the linear excitations in the vacuum cohomology. We find that nilpotence of the BRST operator obstructs most non-linear excitations unless they combine in such way that the vector field induces a metric on $\mathbb{R}^4$, such that the resulting configuration again describes an Einstein space. Thus, the vacuum is continuously connected to the previous solutions described in the first part.

This paper is structured as follows. 

In section \ref{sec:qgr}, we recall the quantization of the space-time graviton based on the BRST quantization of the $\mathcal{N}=4$ superparticle (see \cite{Bonezzi:2018box} and references therein). There we also review how particle excitations are related to background fields for the worldline BRST charge and then deduce the vacuum Einstein equation from nilpotency of the latter.  In section \ref{sec:BI}, we discuss the background independent structure of the $\mathcal{N}=4$ spinning particle and, in particular, determine the BRST cohomology for the vacuum solution with no metric. In addition, we show that an Einstein metric arises necessarily when extending the vacuum fluctuations beyond non-linear order, thus interpolating continuously between the vacuum solution and the Einstein spaces found in the first part of this paper. We then provide a physical interpretation of this deformation and present the conclusions.

\section{Quantization of the graviton}\label{sec:qgr}
Having detected gravitational waves and, indeed, gravitons, it appears natural, in analogy with electromagnetism, to identify the former with a coherent state of gravitons, a massless spin 2 particle state in some Hilbert space expressed in terms of linearized fluctuations of the metric. Having made this association, one needs to take care of the redundancies arising in a description in terms of a metric tensor. There are different ways of doing this. For instance, by a 3+1 splitting of space-time and identifying the two physical degrees of freedom in the spatial part of the metric. Here we will follow the familiar BRST approach where the linear variation of the metric tensor $h_{\mu\nu}$ is combined with an auxiliary vector field, $v_\mu$,  and the diffeomorphism ghost ,$\xi_\mu$, into a BRST multiplet.

\subsection{Spinning particle description}
A convenient (though not unique) way to construct this multiplet is to represent the graviton as an excitation of the relativistic particle  with $\mathcal{N}=4$ worldline supersymmetry. The worldline action of the $\mathcal{N}=4$ super particle is given by 
\begin{equation}\label{eq:wsa}
S=\int \left(p_\mu\dot x^\mu+i\bar\theta^i_\mu\dot\theta^\mu_i-\tfrac{e}{2}\,p^2-i\chi_i\,\bar\theta^i\!\cdot p-i\bar\chi^i\,\theta_i\!\cdot p\right) d\tau\;, \quad i=1,2\,.   
\end{equation}
Here $e(\tau)$ and $\chi_i(\tau)$, $\bar\chi^i(\tau)\,$ are Grassmann even and odd auxiliary fields enforcing the (proper-time) reparametrization constraint $H$, and the local worldline supersymmetry constraints $\bar q$ and $q$ respectively. Let us stress that no space-time supersymmetry is assumed. The spinning worldline is merely a tool introduced here to represent the one-graviton state on some Hilbert space. To see how this comes about we first recall that the quantization of the worldline provides us with the even and odd canonical pairs, 
\begin{equation}
[p_\mu, x^\nu]= \delta_\mu^{\;\nu}\;,\quad\{\bar\theta^i_\mu, \theta_j^\nu\}=\delta^i_j\delta^{\nu}_{\mu} \,,
\label{eq:cp1}
\end{equation}
with all other (anti-) commutators vanishing. We may then build a Fock space whose vacuum is annihilated by $\bar\theta_\mu^i\,$, and a generic state $\ket{\psi}$ in the full Hilbert space described by wave function $\Psi(x^\mu,\theta_i^\mu)$ on which of $p_\mu$ and $\bar\theta^i_{\mu}$ are represented as 
\begin{equation}
 p_\mu=-i\frac{\partial}{\partial x^\mu}\,, \qquad \bar\theta^i_{\mu}=\frac{\partial}{\partial\theta^\mu_i}\,. 
\end{equation} 
The vector space $\mathcal{H}$ obtained in this way contains many more states then needed to describe a graviton. However, following \cite{Bonezzi:2018box}, there is a consistent truncation $\mathcal{H}$ given by the invariant subspace under the  $so(4)$ R-symmetry of the worldline action \eqref{eq:wsa}. 

\subsection{BRST-quantization}
In the BRST quantization of the worldline action \eqref{eq:wsa}, the auxiliary fields multiplying the constraints in \eqref{eq:wsa} give rise to canonical ghost pairs: The reparametrisation ghosts $c$, $b$ and super ghosts $\gamma_i$ and $\beta_i$. The ghost numbers and Grassmann parity of $\left(c,b\right),\left(\gamma_i,\bar\beta^i\right),\left(\bar\gamma^i,\beta_i\right) $ are  $\left( +1 , -1\right), \left(+1 , -1\right),\left( +1, -1\right)$ and  $\left( 1,1\right),\left(0,0\right),\left(0,0\right)$ respectively. They satisfy the canonical (anti-)commutation relations
\begin{equation}
    \{c,b\} = 1 \,, \qquad [\gamma_i,\bar{\beta}^j] = \delta^j_i \,, \qquad [\bar{\gamma}^i,\beta_j] = \delta^i_j\,,
\end{equation}
and, together with \eqref{eq:cp1} generate the algebra $\mathcal{A}$ of the BRST-quantized worldline.  
The space of wave functions is extended accordingly to  $\Psi(x,\theta,c,\beta,\gamma)$ with 
\begin{equation}
  b=\partial_{c}, \qquad \bar\gamma^i=\partial_{\beta_i},\qquad \bar\beta^i=-\partial_{\gamma_i}\,.
\label{eq:GhostDerivatives}
\end{equation} 
Taking into account the action of the  $so(4)$ R-symmetry on the ghost fields, the  $so(4)$ generators are also extended to the ghost sector. The details of all this are described in \cite{Bonezzi:2018box}. The upshot relevant for our purpose is that a general state in our restricted Hilbert space has the form 
\begin{align}
    \tfrac12\,\Phi_{AB}(x)\,Z^A_iZ^B_j\,\epsilon^{ij} \qq{\text{where}} Z^A_i \in \{
    \theta_{i}^{\mu},\gamma_{i},\beta_{i}
    \}.
\label{eq:StatesInHilbertSpace}
\end{align}
Concretely, suppressing antifields,
\begin{align}\label{eq:BRTST_stringfield}
\Psi(x,\theta_i\,c,\gamma_i,\beta_i) =& h_{\mu\nu}(x)\,\theta^\mu_1\theta^\nu_2+\tfrac12\,h(x)\,(\gamma_1\beta_2-\gamma_2\beta_1)-\tfrac{i}{2}\,v_\mu(x)\,(\theta^\mu_1\beta_2-\theta^\mu_2\beta_1)c\nonumber\\
&-\tfrac{i}{2}\,\xi_\mu(x)\,(\theta^\mu_1\beta_2-\theta^\mu_2\beta_1)\,.
\end{align}
Here, $h_{\mu\nu}$ represents the graviton, $h$ its trace, $v_\mu$ an auxiliary vector and $\xi_\mu$ is the diffeomorphism ghost. This explains why the $\mathcal{N}=4$ relativistic point particle provides a representation of the graviton. 

\subsection{BRST-cohomology and physical states}
In the BRST quantization the physical states are in 1-1 correspondence with the cohomology of the BRST operator at ghost number zero. This is also the case for the graviton viewed as an excitation of the spinning worldline whose BRST operator is given by 
\begin{align}
    Q_0 = cH + \gamma_{i} \bar{q}^{i} + \bar{\gamma}^{i}q_{i} + \bar{\gamma}^{i}\gamma_{i}b,
\end{align}
where the subscript zero is to indicate reference to a Minkowski metric background. The equation ${Q_0\Psi=0}$ at ghost number $0$
 then implies
 \begin{equation}
\Box h_{\mu\nu}-\partial_{(\mu}v_{\nu)}=0\;,\quad v_\mu+\partial_\mu h-2\,\partial\cdot h_\mu=0,
\end{equation}
or, after elimination of the auxiliary vector field $v_\mu$,
\begin{equation}
\Box h_{\mu\nu}-2\,\partial_{(\mu}\partial\cdot h_{\nu)}+\partial_\mu\partial_\nu h =0 \;.   
\end{equation}
On the other hand, the diffeomorphism transformation $\delta h_{\mu\nu}=\partial_{(\mu}\xi_{\nu)}$ is encoded in $\delta\Psi=Q_0 \phi\,$ with
\begin{equation}\label{Lambdaparameter}
\phi=\xi_\mu(x)\,(\theta^\mu_1\beta_2-\theta^\mu_2\beta_1)\,.    
\end{equation}
These two equations then establish the equivalence between the BRST-cohomology and the 1-graviton state with the two remaining polarizations. 

\subsection{Coupling to background fields}
\begin{figure}[htb]
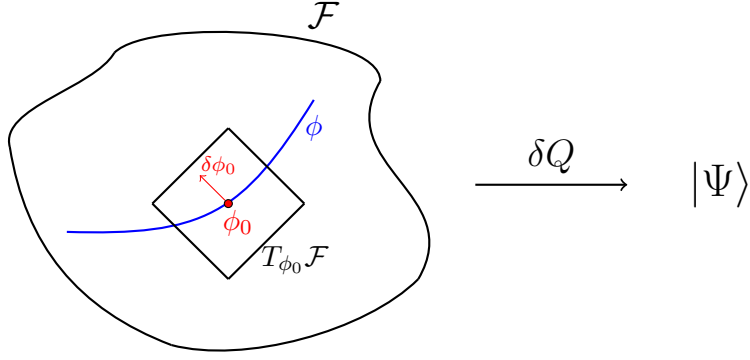

    \centering
    \ctikzfig{FieldSpaceToStateSpace}
    \caption{BRST operator maps from field space to Hilbert space}
    \label{fig:FieldSpace}
\end{figure}

We now consider deformations $Q$ of $Q_0$ such that $Q$ continues to act as a differential on the same vector space $\mathcal{H}$. 
Consider an abstract space $\mathcal{F}$ containing all possible background fields that can couple to the  $\mathcal{N}=4$ superparticle (c.f. Figure \ref{fig:FieldSpace}). Let $\phi \in \mathcal{F}$ be a curve described by one-parameter family of  background fields connected to $\phi_0$ and  
$T_{\phi_{0}}\mathcal{F}$ denote the tangent space at the classical solution $\phi_0$. The tangent space is generally multidimensional and spanned by fluctuations $\delta\phi$ around the classical solution $\phi_{0}$. 
Nilpotent infinitesimal variations $\delta Q$ of a classical background  with respect to the BRST charge $Q_0$ correspond to physical states of our theory. This map is surjective. In particular, it allows for a particle interpretation. 

We should perhaps emphasize, that this ansatz circumvents the introduction of any action, in particular the Einstein-Hilbert action, and would therefore be reproducible by an alien civilization which is ignorant about general relativity. 
 Rather, we derive equations for the background fields that are consistent with the the machinery of BRST quantization. In this way, not only will we reproduce the familiar vacuum Einstein equations but furthermore, using some rather conservative assumptions, one shows that it could not have been any other way. It is a logical consequence from demanding a consistent quantization of our theory. Let us briefly collect some results from \cite{Bonezzi:2018box}: We will not consider the most generic deformation here but instead, following the approach from particle physics, extend $p_\mu$ in \eqref{eq:cp1} to a minimal coupling to geometry, $p_\mu\to \Pi_\mu $, which acts on the wave function \eqref{eq:BRTST_stringfield} as
\begin{align}
    -i\hat\nabla_\mu=-i(\partial_\mu+\omega_{\mu\, ab}\,\theta^a\!\cdot\bar\theta^b)\;,
\end{align}
where $\omega_{\mu\, ab}$ is spin connection and $\theta^a_i$ is related to $\theta^\mu_i$  via the vierbein  $\theta^a_i= e^a_\mu\theta^\mu_i$. In the remainder of this section we simply follow the analysis of  \cite{Bonezzi:2018box}. In particular, one finds that nilpotency of $Q$ implies
\begin{align}\label{eq:qqd}
    0\stackrel{!}{=}\left.\gamma_i\gamma_j\{\bar q^i,\bar q^j\}\right|_{\mathcal{H}}+\left.\bar\gamma^i\bar\gamma^j\{ q_i, q_j\}\right|_{\mathcal{H}}+\left.\gamma_i\bar\gamma^j(\{\bar q^i,q_j\}+\delta^i_{\;j}H)\right|_{\mathcal{H}}\,.
\end{align}
The conditions $\bar\gamma^i\bar\gamma^j\{ q_i, q_j\}=0$ and $\gamma_i\gamma_j\{\bar q^i,\bar q^j\}=0$ are trivially satisfied in ${\mathcal{H}}$, while the last term gives 
\begin{align}\label{Hqq}
  \gamma_i\bar\gamma^j\left.\left( \delta^i_{\;j}(H-\hat{\nabla}^2)-\theta^\mu_i\bar\theta^{\nu\,j}\,(\hat R_{\mu\nu}^\#\;+T_{\mu\nu}^\lambda \partial_\lambda)\right)\right|_{\mathcal{H}} &=0\,,
\end{align}
where $\hat R_{\mu\nu}^\#\; \equiv R_{\mu\nu\lambda\sigma}\,\theta^\lambda\!\cdot\bar\theta^\sigma$ and $T_{\mu\nu}^\lambda$ is the torsion which can not be cancelled by any other term and therefore needs to be set to zero. 

\subsection{Einstein equations from a nilpotent BRST charge}
\label{sec:SpacetimeGeometryFromBRSTCharge}

From our discussion in the previous subsection the BRST charge in a background field is given by 
\begin{align}\label{eq:Qnot}
& Q=c\,H+\bar\gamma^i \theta_i^\mu\Pi_\mu+\gamma^i\bar \theta^{i\mu}\Pi_\mu+\bar\gamma^i\gamma_i\,b\;,
\end{align}
subject to the condition \eqref{Hqq}. On the other hand, in addition to \eqref{eq:qqd} nilpotency of $Q$, implies 
\begin{align}\label{eq:qhi}
    \gamma_i c[\bar \theta^{i\mu}\Pi_\mu,H]= \bar\gamma^i c[ \theta_i^\mu\Pi_\mu,H]=0\,.
\end{align}
As shown in \cite{Bonezzi:2018box} parametrizing $H$ as\footnote{While higher powers of the Riemann tensor or higher order in the derivatives may be consistent with the $\left. Q^2\right|_\mathcal{H}$ they cannot cancel terms that come from $\{\bar q,q\}$.}  
\begin{align}\label{eq:HE}
    H&=\hat{\nabla}^2+\alpha R_{\mu\nu\lambda\sigma}\,\theta^\mu\cdot\bar\theta^{\nu}\theta^\lambda\!\cdot\bar\theta^\sigma\;+F\,,\quad \alpha\in \mathbb{R}\,,
\end{align}
we have 
\begin{equation}\label{second_obstruction}
\begin{split}
\left.[H, \gamma_i \bar \theta^{i\mu}\Pi_\mu]\right|_{\mathcal{H}}  =& \left.2i(1-\alpha)\, \gamma_i \bar \theta^{i\mu}\hat R_{\mu\nu}^\#\hat\nabla^\nu-i\gamma_i \bar \theta^{i\mu}\nabla^\lambda\hat R_{\lambda
\mu}^\#+i(\alpha-1) \gamma_i \bar \theta^{i\mu} R_{\mu\nu}\hat\nabla^\nu\right|_{\mathcal{H}}\\
&+\left.i\alpha \big(2 \nabla^\lambda\hat R_{\lambda
\mu}^\#\gamma\!\cdot\bar\theta^\mu-\gamma_i \bar \theta^{i\mu}\nabla_\mu R_{\nu\lambda}\theta^\nu\!\cdot\bar\theta^\lambda\big)-\gamma_i \bar \theta^{i\mu}\partial_\mu F\right|_{\mathcal{H}}\;,
\end{split}    
\end{equation}
and a similar equation for the second equation in \eqref{eq:qhi}. This fixes $\alpha=1$. Since $T_{\mu\nu}^\lambda=0$, \eqref{Hqq} then simplifies to  

\begin{align}
    \left. \bar\gamma^i\gamma_j
    \left(\delta_i^j
     (R_{\mu\nu\lambda\sigma}\,\theta^\mu\cdot\bar\theta^{\nu}\theta^\lambda\!\cdot\bar\theta^\sigma\;+F)-\theta^\mu_i\bar\theta^{\nu\,j}\,\hat R_{\mu\nu}^\#\;
    \right)
    \right|_{\mathcal{H}} = 0.
\end{align}

Upon substitution into \eqref{Hqq}, the most general state in our restricted Hilbert space (\ref{eq:StatesInHilbertSpace}), which we need to act upon, consists of all six possible\footnote{The number of possible states, allowing for multiple copies of the same element and disregarding different ordering of the elements, is described by the multiset number $\multiset{n}{k} = \frac{(n+k-1)!}{(n-1)!k!}$. In our case we have a total number of $n = |Z_{i}^{A}| = 3 $ elements of which we choose $k=2$ according to \eqref{eq:StatesInHilbertSpace}.} combinations of  $\beta,\gamma  $ and $\theta^{\mu}$. We need to check that each of these states yields zero separately when acted upon by the first term in (\ref{eq:qqd}). However, the presence of the anti-symmetric tensor $\epsilon^{ij}$ in \eqref{eq:StatesInHilbertSpace}  and the prefactor
$\bar{\gamma} = \partial_{\beta} $ in \eqref{Hqq} immediately eliminates all but two states since
\begin{align}
    \epsilon^{ij} \beta_{i}\beta_{j} =  \epsilon^{ij} \gamma_{i}\gamma_{j} = 0, \qq{\text{and}} \pdv{\beta} \theta_{i}^{a}\theta_{i}^{b} = \pdv{\beta}\theta_{i}^{a}\gamma_{j} = 0.
\end{align}
The remaining states on which we still need to evaluate $Q^2$ are given by 
\begin{align}
\Psi = 
\begin{bmatrix}
\ket{\psi_{1}^\alpha} \\
\ket{\psi_{2}} 
\end{bmatrix}
\equiv 
\frac{1}{2}\epsilon^{ij}
\begin{bmatrix}
\theta_{i}^{\alpha}\beta_{j}\Phi_{13} \\
\beta_{i}\gamma_{j} \Phi_{23}
\end{bmatrix}\,.
\label{eq:LeftoverStates}
\end{align}

Let us first consider the state $\ket{\psi_{1}^\alpha}$. Using the algebra satisfied by the fermions \eqref{eq:cp1}, the fact that the anti-ghost acts as a derivative \eqref{eq:GhostDerivatives} and the explicit form of the state \eqref{eq:LeftoverStates} one finds that 
\begin{align}
     \bar\gamma^i\gamma_j
    \left(\delta_i^j
     (R_{\mu\nu\lambda\sigma}\,\theta^\mu\cdot\bar\theta^{\nu}\theta^\lambda\!\cdot\bar\theta^\sigma\;+F)-\theta^\mu_i\bar\theta^{\nu\,j}\,\hat R_{\mu\nu}^\#\;
    \right)\ket{\psi_1^\alpha} = 0,
\end{align}
implies  $R_{\mu\nu} = \lambda(x) g_{\mu\nu}$ with $F(x)=2\lambda(x)$. A detailed derivation of this can be found in appendix \ref{sec:Q2OnPhysicalStates}.

Then, recalling the differential Bianchi identity of the Riemann tensor
\begin{align}
    \nabla_{[\mu}R_{\alpha\beta]\gamma\delta} = 0, 
\end{align}
suitably contracting two indices and subsequently inserting the above result for the Ricci tensor\footnote{and its contracted version, the Ricci scalar $R = g^{\mu\nu}R_{\mu\nu} = g^{\mu\nu}\lambda g_{\mu\nu} = 4 \lambda$.} gives
\begin{align}
  g^{\mu\gamma} g^{\beta\delta}\nabla_{[\mu}R_{\alpha\beta]\gamma\delta} =
  2\nabla_{\mu}R\indices{_\alpha^\mu} - \nabla_{\alpha}R =  -2\nabla_{\alpha}\lambda(x) = 0 \,, 
\end{align}
which fixes our general function to be a constant $ \lambda(x) = \lambda$. The remaining c-ghost term (\ref{second_obstruction}) vanishes on an Einstein manifold since, for $\alpha=1$, the remaining terms, when evaluated on $\mathcal{H}$, are just covariant derivatives of the Ricci tensor. 

The action of \eqref{Hqq} on the second state $\ket{\psi_{2}}$ does not yield any condition. This is because $\ket{\psi_{2}}$ simply commutes with $\bar{\theta}$. Therefore two of three terms in \eqref{Hqq} result in a $\bar{\theta}$ on the right hand side, annihilating the state on the restricted Hilbert space. Furthermore, the last remaining term containing no $\bar{\theta}$ vanishes because of an anti-symmetric contraction resulting from the action of $\bar{\gamma}^{i}$ on the state 

\begin{align}
   \bar{\gamma}^{i}\gamma_{i}F\ket{\psi_{2}} = F \frac{1}{2} \varepsilon^{kl} \pderivative{\beta_{k}}{\beta^{i}}  \gamma_{i} \gamma_{l} \Phi_{23} = F \frac{1}{2} \underbrace{\varepsilon^{il} \gamma_{i} \gamma_{l}}_{=0} \Phi_{23} = 0.
\end{align}

In sum, consistency of the BRST quantization in an arbitrary metric background  implies $ R_{\mu\nu} = \lambda g_{\mu\nu}$.  Thus by simply applying the BRST quantization of the worldline representation of the graviton, without ever having heard of general relativity, our alien civilization could discover the source-less Einstein equations for the spacetime they inhabit. In addition they could postulate the existence of dark energy in form of a cosmological constant. As already mentioned in the introduction the derivation of the non-linear Einstein equations form the linearized ones is not new, having been long ago by different means (\cite{Deser:1969wk} and references therein). The derivation here will turn out to be useful to explore the global properties of the configuration space in the next section.

\section{Background independent structure} 
\label{sec:BI}

As described in \cite{Bonezzi:2018box}, an equivalent parametrization of the phase space of the ${\mathcal{N}=4}$ spinning particle is given by $p_\mu$, $x^\nu$ and four fermionic oscillators  $\theta_{i}^{a}, \bar\theta^{j}_b$, $i,j= 1,2$ with 
\begin{equation}\label{albi1}
[p_\mu,x^\nu]=\delta_\mu^{\;\nu}\;,\quad\{\bar\theta^{i}_a, \theta_{j}^{b}\}=\delta^i_j\,\delta_a^b\;,\quad \{\bar\theta^{ia}, \bar\theta^{jb}\}=0=\{\theta_{i}^{a}, \theta_{j}^{b}\} \;.   
\end{equation}
Here $a,b=0,..,3$ labels a flat non-coordinate basis of $T\mathcal{M}$ with $\theta^\mu_i=e^\mu_a\theta_i^a$. The advantage of the present description is that the space-time metric does not appear in the definition of the oscillator algebra making the latter manifestly  background independent. 

The simplest worldline action for a spinning particle, based on the algebra \eqref{albi1} without any further structure, in particular not assuming a metric structure, is given by
\begin{align}
    S[x,p,\theta,\bar{\theta}]=\int \left(p_\mu\dot x^\mu+i\bar{\theta^i_a}\partial_t\theta_i^a\right)dt=\int p_\mu dx^\mu+i\bar{\theta^i_a} d\theta_i^a\,.
\end{align}
This action is trivially invariant under super- reparametrizations with $x^\mu$, $p_\mu$, $\theta^a_i$ and $\bar{\theta}^{ia}$ simply being invariant, such that  the corresponding constraints vanish. 

Accordingly, the background independent part of the corresponding BRST charge is simply 
\begin{align}\label{eq:q0}
    Q_0&=b\bar{\gamma}^i\gamma_i=-\gamma_i\,\frac{\partial^2}{\partial\beta_i\partial c}\;, 
\end{align}
where the second equality describes the action of $Q_0$ on wave functions $\Psi(x,\theta_i\,c,\gamma_i,\beta_i)\,$.

We should mention that, since all constraints vanish identically, $Q_0\equiv 0$ would also be a consistent choice for $Q_0$ which, however, does not admit any continuous deformations.

\subsection{Cohomology, operator state correspondence and background fields }\label{sec:coho_0}

Let us first identify its cohomology coh$(Q_0,V_0)$ on the reduced vector space $V_0$ spanned by the wave functions $\Psi(x,\theta_i^a\,c,\gamma_i,\beta_i)\,$,
\begin{align}\label{eq:zcoh}
&\Psi(x,\theta_i\,c,\gamma_i,\beta_i)\\ &= h_{ab}(x)\,\theta^a_1\theta^b_2+\tfrac12\,h(x)\,(\gamma_1\beta_2-\gamma_2\beta_1)-\tfrac{i}{2}\,v_a(x)\,(\theta^a_1\beta_2-\theta^a_2\beta_1)c-\tfrac{i}{2}\,\xi_a(x)\,(\theta^a_1\beta_2-\theta^a_2\beta_1)\,.\nonumber
\end{align}
In the cohomology at ghost number zero, this consists of $h_{ab}(x)$ which is symmetric but otherwise unconstrained, a scalar function $h(x)$ which is independent of the trace of $h_{ab}(x)$. The auxiliary field  $v_a=0$ is set to to zero in the cohomology as expected since it is related to the divergence of $h_{ab}(x)$.  

In order to understand the non-linear prolongation of the cohomology, we first need to understand how these states are related to deformations of $Q_0$, by background fields, acting on a reference state in $V_0$. Here we will focus on states of ghost number zero since they have a simple physical interpretation. For this we consider the family of $\mathcal{H}$- preserving deformations\footnote{This may not be a complete set of deformations with this property. }  
\begin{align}\label{eq:delta_Q}
    \delta Q&=
    \bar{\gamma}^i\theta^a_i(E_a +\omega_{abc}\theta^b\cdot\bar{\theta}^c)+ {\gamma}_i\bar{\theta}^{ia}(\bar E_a +\bar\omega_{abc}\theta^b\cdot\bar{\theta}^c)\,, 
\end{align}
where $E_a$, $\bar E_a$, $\omega_{abc}$ and $\bar\omega_{abc}$ may be functions or vector fields, or tensor fields  with values in $\mathcal{A}$, to be specified later. We will also assume $\omega_{abc}=-\omega_{acb}$ and analogously for $\bar\omega_{abc}$. Since we have not endowed $V_0$ with an inner product there is no obvious choice of hermitean conjugation on $Q$ and thus we may  as well consider $E_a$ and $\bar E_a$ as well as $\omega_{abc}$ and $\bar\omega_{abc}$ to be independent. Next we need to specify the reference state. Since $\delta Q$ has ghost number one and we are to produce a physical state in ghost number zero the only available state in in \eqref{eq:zcoh} is 
\begin{align}\label{eq:xiz}
   |\xi>&= \xi_a(x)\,(\theta^a_1\beta_2-\theta^a_2\beta_1),
\end{align}
which we propose to interpret as the ground state wave function of the worldline over which $h_{ab}$ and $h$ describe excitations. We also assume that $\xi^2$ is nowhere vanishing in what follows. Setting  
\begin{align}\label{eq:an1}
   \omega_{abc}&=\frac{1}{2}(h_{ab}k_c-h_{ac}k_b)\;,\quad E_a=\frac{1}{2}h_{ac}k^c\quad\text{with}\quad  k_c=\frac{\xi_c}{\xi^2}
\end{align}
and $\bar E_a=\bar\omega_{abc}=0$ we find that $\delta Q|\xi>$ reproduces the first state of the r.h.s. of \eqref{eq:zcoh}. Similarly, with 
\begin{align}\label{eq:ea1}
  \bar E_a=\frac{1}{2}h(x)\frac{\xi_a}{\xi^2} 
\end{align}
and $ E_a=\omega_{abc}=\bar\omega_{abc}=0$ we reproduce the second term in \eqref{eq:zcoh}. At ghost number zero, this then provides a concrete realization of the surjective map $Q:T\mathcal{F}\to V_0$ in fig \ref{fig:FieldSpace}. This map is clearly not unique. For instance, choosing for $E_a$ a vector field vector field $E_a^\mu \partial_{x^\mu}$ produces, upon action on $|\xi>$, a term of the form  $\Psi=2(E^\mu_a\partial_\mu \xi_b)   \theta^{(a}_1\theta^{b)}_2$. Note also that the map depends on a choice of the wave function $\xi_a$. 

If we furthermore impose the reality condition, $\omega_{abc}=\bar\omega_{abc}$ and $\bar E_a=E_a$ given by \eqref{eq:ea1}, then 
\begin{align}
    \delta Q|\xi>= h_{ab}(x)\,\theta^a_1\theta^b_2+\tfrac12\,h(x)\,(\gamma_1\beta_2-\gamma_2\beta_1)\,,
\end{align}
with $h=h^a_a$.

\subsection{Non-linear field equations}\label{sec:non-lin}
We now turn to the non-linear field equations implied by the nilpotency of $Q(\Phi)$ when acting on a generic state in $V_0$. In this section we will assume reality of the background fields. Recalling \eqref{eq:qqd} we find
\begin{align}\label{Hqq_zm}
     \gamma_i\bar\gamma^j\left.\left( \delta^i_{\;j}(H-\mathcal{E}_a\eta^{ab}\mathcal{E}_b + \eta^{ab}\omega_{ab}^{\;\;\;c}\mathcal{E}_c)-\theta^a_j\bar\theta^{b\,i}\,(\hat{ \mathcal{R}}_{ab}^\#\;+\mathcal{T}_{ab}^c\mathcal{E}_c)\right)\right|_{\mathcal{H}} &= 0,
 \end{align}
where $\mathcal{E}_a=E_a+\omega_{abc}\,\theta^b\!\cdot\bar\theta^c$, $\hat{\mathcal{R}}_{ab}^\#=[\mathcal{E}_a,\mathcal{E}_b]=\mathcal{R}_{abcd}\,\theta^c\!\cdot\bar\theta^d=(\omega_{ac}^{\;\;\;e}\omega_{bed}-\omega_{bc}^{\;\;\;e}\omega_{aed})\,\theta^c\!\cdot\bar\theta^d$ and 
\begin{align}\label{eq:ft}
  \mathcal{T}_{ab}^c \mathcal{E}_c=[E_a,E_b]+(\omega_{abc}-\omega_{bac})\mathcal{E}^c\,,
\end{align}
where $\omega_{abc}=-\omega_{acb}$ was assumed. The first term in \eqref{Hqq} can be cancelled by choosing the (minimal) Hamiltonian as 
\begin{align}
    H=\mathcal{E}_a\eta^{ab}\mathcal{E}_b - \eta^{ab}\omega_{ab}^{\;\;\;c}\mathcal{E}_c\,.
\end{align}
For vanishing or commuting\footnote{This is the case if $E_a$ are functions on $\mathcal{M}$.} $E_a$, vanishing of the "torsion" part then implies $\omega_{abc}=\omega_{bac}$. This, in combination with antisymmetry in $b,c$ implies that $\omega_{abc}$ vanishes identically. Thus, \eqref{Hqq} and \eqref{eq:ft} give an obstruction for the deformation \eqref{eq:delta_Q} at quadratic order. In order to identify further obstructions we then compute

\begin{align}
    \comm{q_{i}}{H} = \theta_{i}^{a}\left(\mathcal{R}_{ac}^{\#}-\mathcal{E}^{d}\, \omega\indices{_{cad}}\right)\left(2\mathcal{E}^{c}-\omega\indices{_e^{ec}}\right)\,.
    \label{eq:qHCommutatorFirstResult}
\end{align}

We still have the freedom to add a non-mimimal coupling 
\begin{equation}\label{eq:HR2}
H\to H + \mathcal{R}_{abcd}\,\theta^a\!\cdot\bar\theta^b\,\theta^c\!\cdot\bar\theta^d\;+F\,,
\end{equation}
where $F$ is an undetermined function. This yields the additional contribution 

\begin{align}
    \comm{\theta_{i}^{c}\, \mathcal{E}_{c}}{\mathcal{R}_{abde}\, \theta^{a}\cdot\bar{\theta}^{b} \theta^{d}\cdot\bar{\theta}^{e}} = 2\theta_{i}^{a}\omega\indices{_{ab}^d}\mathcal{R}_{dc}^{\#} \left(\theta^{b}\cdot\bar{\theta}^{c} + \theta^{c}\cdot\bar{\theta}^{b} \right) + \theta_{i}^{a}(\mathcal{R}_{ac}-2\mathcal{R}_{ac}^{\#})\mathcal{E}^{c},
\end{align}
ultimately resulting in
\begin{multline}
    \comm{q_{i}}{H} =  \theta_{i}^{a}\left(
    \mathcal{E}^d \; \omega\indices{_{cad}}\left(\omega\indices{_{e}^{ec}}-2\mathcal{E}^{c} \right) + \mathcal{R}_{ac}\mathcal{E}^c - \mathcal{R}_{ac}^{\#}\omega\indices{_{e}^{ec}} \right. \\ 
    \left. +\left( \theta^b\!\cdot\bar\theta^d + \theta^d\!\cdot\bar\theta^b \right) \omega\indices{_{ad}^c} \left( 
    \mathcal{R}_{bc} +\mathcal{R}^{\#}_{bc}
    \right) 
    \right).
\label{eq:qHCommutatorSecondResult}
\end{multline}
Thus vanishing \eqref{eq:qHCommutatorSecondResult} implies further obstructions which are, however, of cubic order in the deformation.  Further cubic constraints are obtained form $\left.c\gamma_i[\bar{q}^i,H]\right|_{\mathcal{H}}$. For a detailed computation of \eqref{eq:qHCommutatorFirstResult} and subsequently \eqref{eq:qHCommutatorSecondResult} we refer the reader to the appendix \ref{qHCommutatorComputation}.

To summarize, at linear order in the deformation there are no constraints from nilpotency in agreement with the cohomology identified in section \ref{sec:coho_0}. However, at second order deformations of the form \eqref{eq:delta_Q} are obstructed if we assume that the $E_a$ form a commutative algebra. 

\begin{figure}[htb]
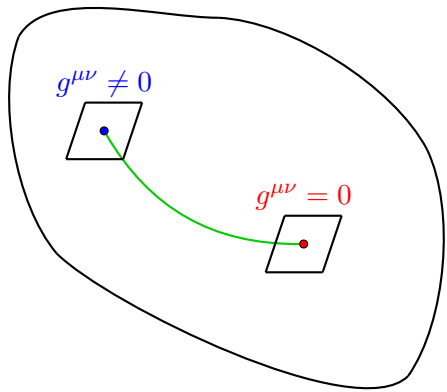

    \centering
    \ctikzfig{VanishingMetric}
    \caption{Flow through field space between the vacuum configuration to a background with metric.}
    \label{fig:VanishingMetric}
\end{figure}

A familiar dynamic arises if we let $E_a$ take values in the tangent bundle, i.e. a collection of vector fields, $E_a^\mu\partial_\mu$. To see this let us return to the torsion equation \eqref{eq:ft}. First we consider $E_a$ as a small perturbation $E_a=O(\epsilon)$, $\epsilon$ small. If $\omega=O(1)$ then \eqref{eq:ft} presents an obstruction to an $O(\epsilon)$ vector field $E_a$. However, we have seen above that $\omega$ itself is obstructed at $O(\omega^3)$. We thus have a linearization instability that prevents a solution for $\omega$ without further background fields beyond second order in perturbation theory. On the other hand if both, $E_a$ and $\omega$ are expanded in powers of $\epsilon$, then the first condition arises from \eqref{eq:ft} at $O(\epsilon^2)$. It can be solved by writing 
\begin{align}
    E_a=\epsilon \hat{E_a}+O(\epsilon^2)
    \,,\quad \omega=\epsilon \hat{\omega}+O(\epsilon^2)\,,
\end{align}
and assuming that the matrix $\hat{E^\mu_a}$ is invertible\footnote{This can be extended to the degenerate case when  $\hat{E^\mu_a}$ is invertible on some subbundle of the tangent bundle.}. Then $\hat\omega_{abc}$ is expressed in terms of $E_a$ through the familiar expression
\begin{align}
    \hat\omega_{abc}=-\frac{1}{2}\left(\hat E^\mu_{[a}\partial_\mu \hat E^\nu_{b]}\right)\hat E_{\nu c}+\frac{1}{2}\left(\hat E^\mu_{[b}\partial_\mu \hat E^\nu_{c]}\right)\hat E^\nu_a-\frac{1}{2}\left(\hat E^\mu_{[c}\partial_\mu \hat E^\nu_{a]}\right)\hat E^\nu_b,
\end{align}
with $\hat E_{\mu c}=\eta_{ca}\hat E_{\mu}^{a}$ where $\hat E_{\mu}^{a}$ is the inverse of $\hat E^{\mu}_{c}$. This procedure can be repeated recursively to higher orders. Having solved the torsion equation \eqref{eq:ft} we still need to satisfy the remaining constraints from $Q^2=0$, in particular \eqref{eq:qqd} and \eqref{eq:qhi} which, as we have seen in the last section imply that either $R_{\mu\nu}=0$ or $R_{\mu\nu}=\frac{1}{2}Fg_{\mu\nu}=\frac{1}{2}F\eta_{ab}E_{\mu}^{a}E_{\nu}^{b}$ with $F$ constant. Thus, we have connected the perturbation around the vacuum configuration without metric or, equivalently, no vector field, to the deformations of Minkowski space-time discussed in the last section, see fig \ref{fig:VanishingMetric}. In particular, a small deformation by a vector field $\epsilon \hat E^{\mu}_{a}=\epsilon\delta^{\mu}_{a}$ corresponds to a large metric space with scale factor $O(1/\epsilon)$. 

\subsection{Physical interpretation}
While setting the derivative terms in $Q$ to zero, or equivalently $e_a^\mu=0=g^{\mu\nu}$, this still leaves the question of what is the physical interpretation of neglecting all but the last term in \eqref{eq:Qnot}. In order to explain this let us recall the consequences of imposing the condition $Q_0\Psi=0$ with $\Psi$ as in \eqref{eq:StatesInHilbertSpace} and $Q_0$ given by \eqref{eq:Qnot}. The term $\gamma_i\bar \theta^{i\mu}\Pi_\mu$ in $Q_0$ implies the equation \cite{Bonezzi:2018box}
\begin{align}
    2\nabla_\mu h^{\mu}_{\;\;\nu}-\partial_\nu h-v_\nu=  2\partial_\mu h^{\mu}_{\;\;\nu}-\Gamma_{\mu\lambda}^\mu h^{\lambda}_{\;\;\nu} +\Gamma_{\mu\nu}^\lambda  h^{\mu}_{\;\;\lambda}-\partial_\nu h-v_\nu=0,
\end{align}
with a similar second equation from the Hamiltonian, $cH$ in $Q_0$. Neglecting the derivative terms in this equation is justified when the background connection  $\Gamma_{\mu\lambda}^\nu$ dominates over the derivatives, which can be interpreted as focussing on gravitons with large wavelength compared to the occupation number of excitations that provide the background. Setting furthermore the background to zero as we did in \eqref{eq:q0}, should then correspond to zero physical (i.e. coordinate invariant) momentum. The cohomology \eqref{eq:zcoh}   
then corresponds to infinitesimal zero momentum excitations $h_{ab}$ which we found not to be obstructed. The fact that the auxiliary field $v_a=0$ in this cohomology is  easily understood given that there is no divergence to compensate. The non-linear analysis in section \ref{sec:non-lin} then shows that these linear excitations can not condense unless a vector field is excited simultaneously inducing, in turn, a metric on $\mathbb{R}^4$. The latter is furthermore constraint to describe an Einstein space.

\section{Conclusion}
In this note, we described a particle physics approach to gravity in which we assumed that a graviton has been detected (admissibly not very realistic) but no other measurement of the large scale structure is available. Quantizing the graviton in the BRST approach realized by a minimal\footnote{Here, minimal means that no higher derivative terms 
are are added to the BRST operator.} spinning worldline, one finds that the non-linear extension of the physical state condition for the graviton implies that the background has to be an Einstein manifold of positive, negative, or vanishing cosmological constant.  

Given the parametrization of the BRST operator in terms of the background gravitational field, we then explored its background independent structure by setting the gravitational field to zero. The corresponding cohomology describes no dynamics but rather, free elementary excitations of a collection of scalar fields and vector fields. Imposing nilpotency of the BRST charge, we then obtain the non-linear extension of the "physical state" condition and found that generic excitations in the cohomology are obstructed at the non-linear level unless they include a vector field that, in turn, induces a metric which has to be Einstein again. Thus the "vacuum" of this model (with no metric) is continuously connected to a Universe realized as an Einstein space.

\acknowledgments
We would like to thank Martín Enríquez Rojo for interesting and helpful discussions.
This work was funded, in parts by the Excellence Cluster Origins of the DFG under Germany’s Excellence Strategy EXC-2094 390783311.

\newpage
\appendix

\section{Evaluate squared BRST charge on physical states}
\label{sec:Q2OnPhysicalStates}

Since $F$ in \eqref{Hqq} acts trivially on the state, we neglect this term for now and focus on the expression

\begin{align}
    \bar{\gamma}^{i}\gamma_{j} \left( 
    \delta\indices{_i^j} \theta^{\mu}\cdot\bar{\theta}^{\nu}\theta^{\lambda}\cdot\bar{\theta}^{\sigma} - \theta^{\mu}_{i}\bar{\theta}^{\nu j}\theta^{\lambda}\cdot\bar{\theta}^{\sigma}
    \right)R_{\mu\nu\lambda\sigma}\ket{\psi_{1}^\alpha}, 
\end{align}

For both terms we need to evaluate the action of a contracted pair $\theta^{\lambda}\cdot\theta^{\sigma}$ on $\ket{\psi_{1}^\alpha}$

\begin{multline}
        \theta^{\lambda}\cdot\bar{\theta}^{\sigma} \ket{\psi_{1}^\alpha} = \theta^{\lambda k}\bar{\theta}^{\sigma}_{k} \frac{1}{2} \varepsilon^{ij}\theta_{i}^{\alpha}\beta_{j}\Phi_{13} = \left. \frac{1}{2} \varepsilon^{ij} \theta^{\lambda k}\acomm{\bar{\theta}^{\sigma}_{k}}{\theta_{i}^{\alpha}} \beta_{j}\Phi_{13} \right|_{\mathcal{H}} \\
        = \left. \frac{1}{2} \varepsilon^{ij} \theta^{\lambda k} \delta_{ki}g^{\sigma\alpha} \beta_{j}\Phi_{13} \right|_{\mathcal{H}}  
    = g^{\sigma\alpha} \left. \frac{1}{2} \varepsilon^{ij} \theta^{\lambda}_{i} \beta_{j}\Phi_{13} \right|_{\mathcal{H}} = g^{\sigma\alpha} \ket{\psi_{1}^\lambda},
    \label{eq:ContractedThetaPairOnState}
\end{multline}

leaving us with

\begin{align}
    \bar{\gamma}^{i}\gamma_{j} \left( 
    \delta\indices{_i^j} \theta^{\mu}\cdot\bar{\theta}^{\nu} - \theta^{\mu}_{i}\bar{\theta}^{\nu j}
    \right)R\indices{_{\mu\nu\lambda}^\alpha}\ket{\psi_{1}^\lambda}. 
    \label{eq:QSquaredHalfEvaluated}
\end{align}

At first glance it seems that these two terms are incompatible with one another. Acting with $\theta^{\mu}_{i}\bar{\theta}^{\nu j}$ on $\ket{\psi_{1}^\lambda}$ results in a state with a different index structure of the form $\varepsilon^{ij} \theta^{\lambda}_{k} \beta_{j}$, i.e. the indices of the anti-symmetric tensor do \emph{not} fully contract with the $\theta\beta$ term. However this can be resolved by  taking the derivative in front of both terms $\bar{\gamma}^{i} = \pderivative{\beta^i}$ into account. 

Let us compute both terms explicitly. Starting with the first term in \eqref{eq:QSquaredHalfEvaluated} and making use of \eqref{eq:ContractedThetaPairOnState}

\begin{multline}
  \bar{\gamma}^{i}\gamma_{j} \delta\indices{_i^j} R\indices{_{\mu\nu\lambda}^\alpha} \theta^{\mu}\cdot\bar{\theta}^{\nu} \ket{\psi_{1}^\lambda} = \bar{\gamma}^{i}\gamma_{i}R\indices{_{\mu\nu\lambda}^\alpha} g^{\nu\lambda} \ket{\psi_{1}^\mu} = -\left. R\indices{_{\mu}^\alpha} \frac{1}{2} \varepsilon^{kl} \theta_{k}^{\mu}\pderivative{\beta_{l}}{\beta^i} \gamma_{i} \Phi_{13}  \right|_{\mathcal{H}} \\
  = -\left. R\indices{_{\mu}^\alpha} \frac{1}{2} \varepsilon^{kl} \theta_{k}^{\mu}\delta_{l}^{i} \gamma_{i} \Phi_{13}  \right|_{\mathcal{H}}  
  = -\left. R\indices{_{\mu}^\alpha} \frac{1}{2} \varepsilon^{ki} \theta_{k}^{\mu} \gamma_{i} \Phi_{13}  \right|_{\mathcal{H}} 
  = -\left. R\indices{_{\mu}^\alpha} \ket{\psi_{1}^\mu} \right|_{\beta\to\gamma}.
\end{multline}

The second term follows from a similar computation

\begin{multline}
    - \bar{\gamma}^{i}\gamma_{j} R\indices{_{\mu\nu\lambda}^\alpha} \theta^{\mu}_{i}\bar{\theta}^{\nu j} \ket{\psi_{1}^\lambda} = \left. - \bar{\gamma}^{i}\gamma_{j} R\indices{_{\mu\nu\lambda}^\alpha}  \frac{1}{2} \varepsilon^{kl} \theta^{\mu}_{i} \acomm{\bar{\theta}^{\nu j}}{\theta_{k}^{\lambda}} \beta_{l}  \Phi_{13}  \right|_{\mathcal{H}} \\ 
    = \left. - \bar{\gamma}^{i}\gamma_{j} R\indices{_{\mu\nu\lambda}^\alpha} g^{\nu\lambda} \frac{1}{2} \varepsilon^{jl} \theta^{\mu}_{i}  \beta_{l}  \Phi_{13}  \right|_{\mathcal{H}} = \left. - R\indices{_{\mu}^\alpha}  \frac{1}{2} \varepsilon^{lj} \theta^{\mu}_{i} \pderivative{\beta_{l}}{\beta^{i}}  \gamma_{j}   \Phi_{13}  \right|_{\mathcal{H}} = \left.  -R\indices{_{\mu}^\alpha}  \frac{1}{2} \varepsilon^{ij} \theta^{\mu}_{i}  \gamma_{j}   \Phi_{13}  \right|_{\mathcal{H}} \\ 
    = \left. -R\indices{_{\mu}^\alpha} \ket{\psi_{1}^\mu} \right|_{\beta\to\gamma}.
\end{multline}

Lastly, consider the $F$ term we neglected thus far

\begin{align}
    \bar{\gamma}^{i}\gamma_{j} \delta\indices{_i^j} F \ket{\psi_{1}^\alpha} = \left. F g^{\alpha}_{\mu}  \ket{\psi_{1}^\mu} \right|_{\beta\to\gamma}.
\end{align}

Putting everything together 

\begin{align}
    \left. \left(  Fg\indices{_{\mu}^\alpha}  - 2R\indices{_{\mu}^\alpha} \right) \ket{\psi_{1}^\mu} \right|_{\beta\to\gamma} = 0.
\end{align}

\section{Cubic spin-connection obstruction}
\label{qHCommutatorComputation}

In this appendix we gather some of the computations needed for evaluating the obstructions to the deformation of the BRST charge $Q$, given by $\comm{q}{H}$. Recall that the full Hamiltonian including the non-minimal coupling term reads
\begin{align}
    H=\mathcal{E}_a\eta^{ab}\mathcal{E}_b - \eta^{ab}\omega\indices{_{ab}^c} \mathcal{E}_c +  \mathcal{R}_{abcd}\,\theta^a\!\cdot\bar\theta^b\,\theta^c\!\cdot\bar\theta^d +  F. 
\end{align}
The task of computing the obstructions can therefore be broken down into evaluating three commutators 
\begin{align}
     \comm{q_{i}}{H} = \underbrace{\comm{\theta_{i}^{c}\mathcal{E}_c}{\mathcal{E}_a\eta^{ab}\mathcal{E}_b}}_{\#1}  \underbrace{-\comm{\theta_{i}^{c}\mathcal{E}_c}{ \eta^{ab}\omega\indices{_{ab}^d} \mathcal{E}_d}}_{\#2} + \underbrace{\comm{\theta_{i}^{c}\mathcal{E}_c}{\mathcal{R}_{abde}\,\theta^a\!\cdot\bar\theta^b\,\theta^d\!\cdot\bar\theta^e}}_{\#3}.
\end{align}
To this end, we only need to make use of some basic relations gathered below for convenience. We assume a spin connection which is anti-symmetric in its last two indices    
\begin{align}
    \omega_{abc} = -\omega_{acb},
\end{align}
recall the algebra obeyed by our fermion fields 
\begin{align}
    \acomm{\theta_{i}^{a}}{\bar{\theta}^{bj}} = \eta^{ab}\delta^{j}_{i},
\end{align}
as well as the symmetries of the Riemann tensor
\begin{align}
    \mathcal{R}_{abcd} =  -\mathcal{R}_{bacd} =  -\mathcal{R}_{abdc} = \mathcal{R}_{cdab},
\end{align}
and its contracted form, the Ricci tensor 

\begin{align}
     \mathcal{R}_{ab} = \mathcal{R}\indices{^c_{acb}} = \eta^{cd}\mathcal{R}\indices{_{cadb}}, \qquad \mathcal{R}_{ab} = \mathcal{R}_{ba}.
\end{align}

For the sake of clarity, we list all (anti-)commutators needed below in order of increasing complexity. Additional computational steps are explicitly shown if deemed helpful. 
We start with some basic commutators

\begin{align}
    \comm{\theta_{i}^{a}}{\theta^{b}\cdot\bar{\theta}^{c}} = -\eta^{ac}\theta_{i}^{b}, \qquad   \comm{\bar{\theta}_{i}^{a}}{\theta^{b}\cdot\bar{\theta}^{c}} = \eta^{ac}\bar{\theta}_{i}^{b},
\end{align}
\begin{align}
    \comm{\theta^{a}\cdot\bar{\theta}^{b}}{\theta^{c}\cdot\bar{\theta}^{d}} = \theta^{a}\cdot\bar{\theta}^{d}\eta^{bc} - \theta^{c}\cdot\bar{\theta}^{b}\eta^{ad},
\end{align}
\begin{align}
    \comm{\theta_{i}^{a}}{\mathcal{E}_{b}} = \omega\indices{_b^a_c}\theta_{i}^{c}, \qquad  \comm{\bar{\theta}_{i}^{a}}{\mathcal{E}_{b}} =  \omega\indices{_b^a_c}\bar{\theta}_{i}^{c},
 \end{align}
\begin{align}
    \comm{\mathcal{E}_{a}}{\mathcal{E}_{b}} = \omega_{acd}\omega_{bef}\comm{\theta^{c}\cdot\bar{\theta}^{d}}{\theta^{e}\cdot\bar{\theta}^{f}} = \mathcal{R}_{abde}\theta^{d}\cdot\bar{\theta}^{e} = \mathcal{R}_{ab}^{\#},
\end{align}
which can be used to compute the more complex commutators
\begin{align}
    \comm{\mathcal{E}_{c}}{\mathcal{E}_{a}\mathcal{E}_{b}} = \mathcal{E}_{a}\mathcal{R}_{cb}^{\#} + \mathcal{R}_{ca}^{\#}\mathcal{E}_{b},
\end{align}

\begin{align}
    \comm{\theta_{i}^{c}}{\mathcal{E}_{a}\mathcal{E}_{b}} = - \left( 
    \mathcal{E}_{a}\omega\indices{_{bd}^c}\theta^{d}_{i} + \omega\indices{_{ad}^c}\theta^{d}_{i}\mathcal{E}_{b}
    \right),
\end{align}
which, in fact, already suffice for both terms $\#1$ and $\#2$. However, term $\#3$ requires some more involved commutators
\begin{align}
    \comm{\mathcal{E}_{a}}{\theta^{b}\cdot\bar{\theta}^{c}} = \omega\indices{_{ad}^b}\theta^{d}\cdot\bar{\theta}^{c} + \omega\indices{_{ad}^c}\theta^{b}\cdot\bar{\theta}^{d},
\end{align}
\begin{align}
    \comm{\theta_{i}^c}{\theta^a\! \cdot \bar{\theta}^b \theta^d\! \cdot \bar{\theta}^e} &= \theta^a\! \cdot \bar{\theta}^b \comm{\theta_{i}^c}{\theta^d\! \cdot \bar{\theta}^e} + \comm{\theta_{i}^c}{\theta^a\! \cdot \bar{\theta}^b} \theta^d\! \cdot \bar{\theta}^e \nonumber \\
    &= -\left(
    \theta^a\! \cdot \bar{\theta}^b \theta_{i}^d \eta^{ce} + \theta_{i}^a \theta^d\! \cdot \bar{\theta}^e \eta^{cb} 
    \right) \nonumber \\ 
    &= -\left(
    \theta_{i}^a \eta^{ce}\eta^{bd} + \theta_{i}^d \theta^a\! \cdot \bar{\theta}^b \eta^{ce}  + \theta_{i}^a \theta^d\! \cdot \bar{\theta}^e \eta^{cb}
    \right).
    \label{eq:CommFourThetaTerm}
\end{align}
For the sake of clarity, we will combine \eqref{eq:CommFourThetaTerm} with the terms as they will ultimately appear in the computation of $\#3$ 
\begin{align}
    &\phantom{=}\mathcal{R}_{abde} \comm{\theta_{i}^c}{\theta^a\! \cdot \bar{\theta}^b \theta^d\! \cdot \bar{\theta}^e} \mathcal{E}_{c} \nonumber \\
    &= -\mathcal{R}_{abde} \left(
    \theta_{i}^a \eta^{ce}\eta^{bd} + \theta_{i}^d \theta^a\! \cdot \bar{\theta}^b \eta^{ce}  + \theta_{i}^a \theta^d\! \cdot \bar{\theta}^e \eta^{cb}
    \right) \mathcal{E}_{c} \nonumber \\ 
    &= -\theta_{i}^a \mathcal{R}_{abde} \left(
    \eta^{ce}\eta^{bd} + 2 \theta^d\! \cdot \bar{\theta}^e \eta^{cb} \right) \mathcal{E}_{c}  \nonumber \\ 
    &= \theta_{i}^a \left(
    \mathcal{R}_{ac}-2\mathcal{R}_{ac}^{\#} 
    \right) \mathcal{E}^{c}, 
\end{align}
\begin{align}
    &\phantom{=}\comm{\mathcal{E}_{c}}{\theta^a\! \cdot \bar{\theta}^b \theta^d\! \cdot \bar{\theta}^e} \nonumber \\
    &= \theta^a\! \cdot \bar{\theta}^b \comm{\mathcal{E}_{c}}{\theta^d\! \cdot \bar{\theta}^e} + \comm{\mathcal{E}_{c}}{\theta^a\! \cdot \bar{\theta}^b } \theta^d\! \cdot \bar{\theta}^e \nonumber \\ 
    &= \theta^a\! \cdot \bar{\theta}^b \left( 
    \omega\indices{_{cf}^d} \; \theta^f\! \cdot \bar{\theta}^e + \omega\indices{_{cf}^e} \; \theta^d\! \cdot \bar{\theta}^f 
    \right)
    + 
    \left( 
    \omega\indices{_{cfa}} \; \theta^f\! \cdot \bar{\theta}^b + \omega\indices{_{cfb}} \; \theta^a\! \cdot \bar{\theta}^f 
    \right) \theta^d\! \cdot \bar{\theta}^e \nonumber \\ 
    &= \omega\indices{_{cf}^d} \; \theta^a\! \cdot \bar{\theta}^b \theta^f\! \cdot \bar{\theta}^e 
    +  \omega\indices{_{cf}^e} \; \theta^a\! \cdot \bar{\theta}^b \theta^d\! \cdot \bar{\theta}^f 
    + \omega\indices{_{cf}^a} \; \theta^f\! \cdot \bar{\theta}^b \theta^d\! \cdot \bar{\theta}^e 
    + \omega\indices{_{cf}^b} \; \theta^a\! \cdot \bar{\theta}^f \theta^d\! \cdot \bar{\theta}^e,
    \label{eq:CommEFourTheta}
 \end{align}
\begin{align}
    \acomm{\mathcal{R}_{de}^{\#}}{\theta^f\! \cdot \bar{\theta}^e} &= \mathcal{R}_{deab}\; \acomm{\theta^a\! \cdot \bar{\theta}^b}{\theta^f\! \cdot \bar{\theta}^e} \nonumber \\ 
    &= \mathcal{R}_{deab}\; \left(
    \comm{\theta^a\! \cdot \bar{\theta}^b}{\theta^f\! \cdot \bar{\theta}^e} + 2\theta^f\! \cdot \bar{\theta}^e \theta^a\! \cdot \bar{\theta}^b
    \right) \nonumber \\ 
    &= \mathcal{R}_{deab}\; \left(
    \theta^{a}\cdot\bar{\theta}^{f}\eta^{be} - \theta^{f}\cdot\bar{\theta}^{b}\eta^{ae}     + 2\theta^f\! \cdot \bar{\theta}^e \theta^a\! \cdot \bar{\theta}^b
    \right) \nonumber \\ 
    &= 
    \mathcal{R}_{da}\; \theta^{a}\cdot\bar{\theta}^{f} +  \mathcal{R}_{db}\;\theta^{f}\cdot\bar{\theta}^{b} + 2\theta^f\! \cdot \bar{\theta}^e \mathcal{R}_{de}^{\#} \nonumber \\ 
    &= \mathcal{R}_{da}\; \left(\theta^{a}\cdot\bar{\theta}^{f} + \theta^{f}\cdot\bar{\theta}^{a}\right) + 2\theta^f\! \cdot \bar{\theta}^e \mathcal{R}_{de}^{\#},
\end{align}
\begin{align}
    \acomm{\mathcal{R}_{de}^{\#}}{\theta^e\! \cdot \bar{\theta}^f} &= \mathcal{R}_{deab}\; \acomm{\theta^a\! \cdot \bar{\theta}^b}{\theta^f\! \cdot \bar{\theta}^e} \nonumber \\ 
    &= \mathcal{R}_{deab}\; \left(
    \theta^{a}\cdot\bar{\theta}^{e}\eta^{bf} - \theta^{e}\cdot\bar{\theta}^{b}\eta^{af}     + 2\theta^e\! \cdot \bar{\theta}^f \theta^a\! \cdot \bar{\theta}^b
    \right) \nonumber \\ 
    &= 
    \mathcal{R}\indices{_{dea}^f}\; \left( \theta^{a}\cdot\bar{\theta}^{e} + \theta^{e}\cdot\bar{\theta}^{a} \right) + 2\theta^e\! \cdot \bar{\theta}^f \mathcal{R}_{de}^{\#}.
    \label{eq:AntiCommRHashtag2}
\end{align}
Using \eqref{eq:CommEFourTheta}, factoring out the $\theta$'s by making generous use of index re-labeling 
\begin{align}
    &\phantom{=}\theta_{i}^{c}\mathcal{R}_{abde} \comm{\mathcal{E}_c}{\theta^a\!\cdot\bar\theta^b\,\theta^d\!\cdot\bar\theta^e} \nonumber \\ 
    &= \theta_{i}^{c}\left(
    \omega\indices{_{cf}^d}\;\mathcal{R}_{abde}
    + \omega\indices{_{ce}^d}\;\mathcal{R}_{abfd}
    + \omega\indices{_{ca}^d}\;\mathcal{R}_{dbfe}
    + \omega\indices{_{cb}^d}\;\mathcal{R}_{adfe}
    \right)\theta^a\!\cdot\bar\theta^b\,\theta^f\!\cdot\bar\theta^e \nonumber \\ 
    &= \theta_{i}^{c}\left(
    \omega\indices{_{cf}^d}\;\mathcal{R}^{\#}_{de}
    + \omega\indices{_{ce}^d}\;\mathcal{R}^{\#}_{fd}
    \right)\theta^f\!\cdot\bar\theta^e 
    +\theta_{i}^{c}\theta^a\!\cdot\bar\theta^b \left(
     \omega\indices{_{ca}^d}\;\mathcal{R}^{\#}_{db}
    + \omega\indices{_{cb}^d}\;\mathcal{R}^{\#}_{ad}
    \right) \nonumber \\ 
    &= \theta_{i}^{c} \acomm{\left(
    \omega\indices{_{cf}^d}\;\mathcal{R}^{\#}_{de}
    + \omega\indices{_{ce}^d}\;\mathcal{R}^{\#}_{fd}
    \right)}{\theta^f\!\cdot\bar\theta^e} \nonumber \\ 
    &= \theta_{i}^{c}\omega\indices{_{cf}^d}\left(
    \acomm{\mathcal{R}^{\#}_{de}}{\theta^f\!\cdot\bar\theta^e}
    + \acomm{\mathcal{R}^{\#}_{de}}{\theta^e\!\cdot\bar\theta^f}
    \right) \nonumber \\ 
    &= \theta_{i}^{c}\left( \theta^a\!\cdot\bar\theta^f + \theta^f\!\cdot\bar\theta^a \right) \omega\indices{_{cf}^d} \left( 
    \mathcal{R}_{ad} +\mathcal{R}^{\#}_{ad}
    \right),
 \end{align} 
where from the penultimate line to the result we re-labeled some indices again and used that the term including the full Riemann tensor $\mathcal{R}\indices{_{dea}^f}$ from \eqref{eq:AntiCommRHashtag2} vanishes identically because 
\begin{align}
    \theta_{i}^{c}\left( \theta^a\!\cdot\bar\theta^f + \theta^f\!\cdot\bar\theta^a \right)\omega\indices{_c^d_e}\; \mathcal{R}\indices{_d_f^e_a}
    &= \theta_{i}^{c}\left( \theta^a\!\cdot\bar\theta^f + \theta^f\!\cdot\bar\theta^a \right)\omega\indices{_c^e_d}\; \mathcal{R}\indices{_e_f^d_a} \nonumber \\
    &= \theta_{i}^{c}\left( \theta^a\!\cdot\bar\theta^f + \theta^f\!\cdot\bar\theta^a \right)(-\omega\indices{_c^d_e})\; \mathcal{R}\indices{_d_a^e_f} \nonumber \\
    &= \theta_{i}^{c}\left( \theta^f\!\cdot\bar\theta^a + \theta^a\!\cdot\bar\theta^f \right)(-\omega\indices{_c^d_e})\; \mathcal{R}\indices{_d_f^e_a} \nonumber \\
     &= - \theta_{i}^{c}\left( \theta^a\!\cdot\bar\theta^f + \theta^f\!\cdot\bar\theta^a \right)\omega\indices{_c^d_e}\; \mathcal{R}\indices{_d_f^e_a} \nonumber \\
     &\Rightarrow \theta_{i}^{c}\left( \theta^a\!\cdot\bar\theta^f + \theta^f\!\cdot\bar\theta^a \right)\omega\indices{_c^d_e}\; \mathcal{R}\indices{_d_f^e_a} = 0.
\end{align}
These are all the commutators we need. We can now straight forwardly write  
\begin{multline}
    \#1 = \comm{\theta_{i}^{c}\mathcal{E}_c}{\mathcal{E}_a\eta^{ab}\mathcal{E}_b} = \eta^{ab} \comm{\theta_{i}^{c}\mathcal{E}_c}{\mathcal{E}_a\mathcal{E}_b} 
    = \eta^{ab} \left( 
    \theta_{i}^{c}\comm{\mathcal{E}_c}{\mathcal{E}_a\mathcal{E}_b} + \comm{\theta_{i}^{c}}{\mathcal{E}_a\mathcal{E}_b}\mathcal{E}_c
    \right) \\
    = 2\theta_{i}^{c}\mathcal{E}^{a}\mathcal{R}_{ca}^{\#} -2 \theta_{i}^{d}\mathcal{E}^{a}\mathcal{E}_{c}\;\omega\indices{_{ad}^{c}} 
    = 2\theta_{i}^{a}\mathcal{E}^{c}\left(
    \mathcal{R}_{ac}^{\#} - \mathcal{E}_{d}\;\omega\indices{_{ca}^{d}}
    \right),
\end{multline}
and
\begin{multline}
    \#2 = -\comm{\theta_{i}^{c}\mathcal{E}_c}{ \eta^{ab}\omega\indices{_{ab}^d} \mathcal{E}_d} = -\eta^{ab}\omega\indices{_{ab}^d} \; \comm{\theta_{i}^{c}\mathcal{E}_c}{ \mathcal{E}_d} = -\eta^{ab}\omega\indices{_{ab}^d} \; \left( 
    \theta_{i}^{c}\comm{\mathcal{E}_c}{\mathcal{E}_d} + \comm{\theta_{i}^c}{\mathcal{E}_d}\mathcal{E}_c
    \right)  \\ 
    = -\eta^{ab}\omega\indices{_{ab}^d} \; \left( 
    \theta_{i}^{c}\mathcal{R}_{cd}^{\#} + \omega\indices{_d^c_e}\theta_{i}^{e} \mathcal{E}_c
    \right) = - \omega\indices{_f^{fc}} \; \theta_{i}^{a} \left(
    \mathcal{R}_{ac}^{\#} - \mathcal{E}_d \; \omega\indices{_{ca}^d}
    \right).
\end{multline}
Adding both of these results reproduces \eqref{eq:qHCommutatorFirstResult} as claimed
\begin{align}
    \#1 + \#2 = \theta_{i}^{a}\left(\mathcal{R}_{ac}^{\#}-\mathcal{E}^{d}\, \omega\indices{_{cad}}\right)\left(2\mathcal{E}^{c}-\omega\indices{_e^{ec}}\right).
\end{align}

The term resulting from non-minimal coupling takes the form
\begin{align}
    \#3 &= \comm{\theta_{i}^{c}\mathcal{E}_c}{\mathcal{R}_{abde}\,\theta^a\!\cdot\bar\theta^b\,\theta^d\!\cdot\bar\theta^e} \nonumber \\
    &= \theta_{i}^{c}\mathcal{R}_{abde} \comm{\mathcal{E}_c}{\theta^a\!\cdot\bar\theta^b\,\theta^d\!\cdot\bar\theta^e} + \mathcal{R}_{abde}\comm{\theta_{i}^{c}}{\theta^a\!\cdot\bar\theta^b\,\theta^d\!\cdot\bar\theta^e}\mathcal{E}_c \nonumber \\ 
    &= \theta_{i}^{c}\left( \theta^a\!\cdot\bar\theta^f + \theta^f\!\cdot\bar\theta^a \right) \omega\indices{_{cf}^d} \left( 
    \mathcal{R}_{ad} +\mathcal{R}^{\#}_{ad}
    \right) 
    + \theta_{i}^a \left(
    \mathcal{R}_{ac}-2\mathcal{R}_{ac}^{\#} 
    \right) \mathcal{E}^{c}. 
\end{align}
Let us note that we have not restricted on the Hilbert space $\mathcal{H}$ at any point to obtain this result. Putting it all together and simplifying, we finally obtain 

\begin{multline}
    \comm{q_{i}}{H} =  \theta_{i}^{a}\left(
    \mathcal{E}^d \; \omega\indices{_{cad}}\left(\omega\indices{_{e}^{ec}}-2\mathcal{E}^{c} \right) + \mathcal{R}_{ac}\mathcal{E}^c - \mathcal{R}_{ac}^{\#}\omega\indices{_{e}^{ec}} \right. \\ 
    \left. +\left( \theta^b\!\cdot\bar\theta^d + \theta^d\!\cdot\bar\theta^b \right) \omega\indices{_{ad}^c} \left( 
    \mathcal{R}_{bc} +\mathcal{R}^{\#}_{bc}
    \right) 
    \right).
\end{multline}

\newpage
\bibliographystyle{JHEP}
\providecommand{\href}[2]{#2}\begingroup\raggedright\endgroup

\end{document}

%% file: sample.tikzstyles
\tikzstyle{Green Node}=[fill=green, draw=black, shape=circle, inner sep=0pt, minimum size=3pt]
\tikzstyle{Red Node}=[fill=red, draw=black, shape=circle, inner sep=0pt, minimum size=3pt]
\tikzstyle{Blue Node}=[fill=blue, draw=black, shape=circle, inner sep=0pt, minimum size=3pt]

\tikzstyle{thick black edge}=[-, draw=black, fill=none, thick]
\tikzstyle{Very thick black edge}=[-, very thick, fill=none, draw=black]
\tikzstyle{Thick Blue Edge}=[-, draw=blue, thick]
\tikzstyle{Thick Orange Edge}=[-, draw={rgb,255: red,255; green,128; blue,0}, thick]
\tikzstyle{Thick Red Edge}=[-, draw=red, thick]
\tikzstyle{Thick Green Edge}=[-, draw=green, thick]
\tikzstyle{Black Arrow}=[->, draw=black]
\tikzstyle{Red Arrow}=[draw=red, ->]
\tikzstyle{Thick Black Arrow}=[draw=black, ->, thick]
\tikzstyle{BlackDashedEdge}=[-, dashed, draw=black]
\tikzstyle{DashedRedEdge}=[-, draw=red, dashed]
\tikzstyle{Darker Green Edge}=[-, draw={rgb,255: red,0; green,204; blue,0}, thick]